\documentclass[conference]{IEEEtran}
\IEEEoverridecommandlockouts
\usepackage{cite}
\usepackage{amsmath,amssymb,amsfonts}
\usepackage{algpseudocode}
\usepackage{algorithm}
\usepackage{graphicx}
\usepackage{textcomp}
\usepackage{xcolor}
\usepackage{hyperref}
\usepackage[nolist]{acronym}

\begin{acronym}
    \acro{HPA}{high power amplifier}
    \acro{MIMO}{multiple-input, multiple-output}
    \acro{PAPR}{peak-to-average power ratio}
    \acro{AWGN}{additive white Gaussian noise}
    \acro{KPI}{key performance indicator}
    \acro{KPIs}{key performance indicators}  
    \acro{D2D}{device-to-device}
    \acro{ISAC}{integrated sensing and communication}
    \acro{DFRC}{dual-functional radar and communication}
    \acro{AoA}{angle of arrival}
    \acro{ToA}{time of arrival}
    \acro{EVM}{error vector magnitude}
    \acro{CPI}{coherent processing interval}
    \acro{eMBB}{enhanced mobile broadband}
    \acro{URLLC}{ultra-reliable low latency communications}
    \acro{mMTC}{massive machine type communications}
	\acro{QCQP}{quadratic constrained quadratic programming}
	\acro{BnB}{branch and bound}
	\acro{SER}{symbol error rate}
	\acro{LFM}{linear frequency modulation}
	\acro{ADMM}{alternating direction method of multipliers}
	\acro{BS}{base station}
	\acro{LoS}{line-of-sight}
	\acro{CFO}{carrier frequency offset}
	\acro{UL}{uplink}
	\acro{DL}{downlink}
	\acro{ULA}{uniform linear antenna array}
	\acro{MOO}{multi-objective optimization}
	\acro{IBO}{input back-off}
	\acro{QPSK}{quadrature phase shift keying}
	\acro{CCDF}{complementary cumulative distribution function}
	\acro{OFDM}{orthogonal frequency-division multiplexing}
	\acro{QAM}{quadrature amplitude modulation}
	\acro{IRS}{intelligent reflecting surface}
\end{acronym}

\DeclareMathOperator{\SNR}{SNR}
\DeclareMathOperator{\SINR}{SINR}
\DeclareMathOperator{\MUI}{MUI}
\DeclareMathOperator{\PAPR}{PAPR}
\DeclareMathOperator{\comm}{comm}
\DeclareMathOperator{\ve}{vec}
\DeclareMathOperator{\Real}{Re}
\DeclareMathOperator{\Imag}{Im}
\DeclareMathOperator{\CCDF}{CCDF}

\usepackage{fancyhdr}
\fancypagestyle{firststyle}{
	\fancyhf{}
	\fancyhead[L]{A. Bazzi and M. Chafii, ``Designing Waveforms with Adjustable PAPR for Integrated Sensing and Communication'', accepted in \textit{IEEE International Conference on Communications} (ICC), June, 2025.}

}

\def\BibTeX{{\rm B\kern-.05em{\sc i\kern-.025em b}\kern-.08em
    T\kern-.1667em\lower.7ex\hbox{E}\kern-.125emX}}
\begin{document}

\title{Designing Waveforms with Adjustable PAPR for Integrated Sensing and Communication}

%\author{\IEEEauthorblockN{1\textsuperscript{st} Ahmad Bazzi}
%\IEEEauthorblockA{\textit{Engineering Division} \\
%\textit{New York University Abu Dhabi}\\
%Abu Dhabi, UAE \\
%ahmad.bazzi@nyu.edu}
%\and
%\IEEEauthorblockN{2\textsuperscript{nd} Marwa Chafii}
%\IEEEauthorblockA{\textit{Engineering Division} \\
%\textit{New York University Abu Dhabi}\\
%Abu Dhabi, UAE \\}
%\IEEEauthorblockA{\textit{NYU WIRELESS} \\
%\textit{NYU Tandon School of Engineering}\\
%Brooklyn, NY \\
%marwa.chafii@nyu.edu}
%}

\author{\IEEEauthorblockN{Ahmad Bazzi\IEEEauthorrefmark{1}\IEEEauthorrefmark{2} and Marwa Chafii\IEEEauthorrefmark{1}\IEEEauthorrefmark{2}}
\IEEEauthorblockA{\IEEEauthorrefmark{1}Engineering Division, New York University (NYU), Abu Dhabi, UAE.}
	\IEEEauthorblockA{\IEEEauthorrefmark{2}NYU WIRELESS, NYU Tandon School of Engineering, New York, USA}}

\maketitle
\thispagestyle{firststyle}

\begin{abstract}
This paper presents a new optimization framework dedicated for integrated sensing and communication (ISAC) waveform design. In particular, the problem aims at maximizing the total achievable sum-rate, through multi-user interference minimization, while preserving a certain level of similarity to a given desired radar waveform. Aiming towards feasible and practical PHY architectures, we also offer the flexibility of tuning the peak-to-average power ratio to a desired level. Towards this design, a non-convex optimization problem is formulated, and an alternating direction method of multipliers based solution is derived to converge towards the superiority of the final ISAC waveform. Finally, simulation results validate the proposed ISAC waveform design, as compared to state-of-the-art solutions.  
\end{abstract}
%We present a novel approach to the problem of dual-functional radar and communication (DFRC) waveform design with adjustable peak-to-average power ratio (PAPR), while minimizing the multi-user communication interference and maintaining a similarity constraint towards a radar chirp signal. The approach is applicable to generic radar chirp signals and for different constellation sizes. We formulate the waveform design problem as a non convex optimization problem. As a solution, we adopt the alternating direction method of multipliers (ADMM), hence iterating towards a stable waveform for both radar and communication purposes. Additionally, we prove convergence of the proposed iterative waveform design and demonstrate its superior performance by computer simulations, in comparison to state-of-the-art radar-communication waveform designs.


\section{Introduction}
Researchers have started to explore higher frequency ranges, spanning  $300$GHz to $3$THz \cite{8732419} to realize new technologies.
In fact, one of the key ingredients of next generation wireless communication networks 6G mobile radio is \ac{ISAC} \cite{cheng2022channel}. Indeed, \ac{ISAC} strengthens high-quality wireless connectivity with support to high data transmission rates and accurate sensing features \cite{10373185}. Therefore, diligence during the waveform design process is required to achieve such robustness in sensing, as well as high quality of communication \cite{xiao2022waveform}. Owing to its dual nature, \ac{ISAC} architectures come with a handful of merits, such as hardware efficiency and shared resources \cite{chafii2022ten}.

 \ac{ISAC} can be split into three broad categories, where the first aims at a \textit{joint design}, offering trade-offs between sensing and communications. For instance, \cite{bazzi2022outage} derives beamforming matrices for \ac{ISAC} systems, in situations of imperfect channel state information, while \cite{xu2022robust} considers a joint design for variable length time snapshots. Another category is \textit{communication-centric} \ac{ISAC}, where one simply performs sensing using a known communication waveform, such as \ac{OFDM}. The priority in this case is communications, and sensing comes as an add-on. This is interesting in applications where the communication infrastructure is established and follows a certain standard, ex. IEEE 802.11aj. For example, the novel design in \cite{10417003} makes use of already-existing communication signals to perform sensing in a hybrid fashion and \cite{bazzi2022ris} utilizes an \ac{IRS} to significantly enhance sensing performances. Furthermore, \textit{radar-centric} \ac{ISAC}'s main goal is to embed communication information onto a radar waveform. See \cite{9828505} for chirp waveforms conveying communication information. Now, we state some state-of-the-art methods related to \ac{ISAC} waveform designs.
Also, the work in \cite{9593174} discusses a possible \ac{ISAC} demonstration testbed for millimeter-wave applications.
 
%% Existing work
The work in \cite{hu2022low} focuses on spatial beamforming for the radar sub-system, whereas our work considers similarity constraints relative to a chirp with good auto-correlation properties, such as good ambiguity function properties, which further empower good \ac{ToA} estimations \cite{bazzi2015efficient} via the received radar echo. Moreover, the work in \cite{hu2022low} examines waveforms for \ac{MIMO}-\ac{OFDM} intended for \ac{DFRC} systems. In addition, the problem in \cite{hu2022low} considers a weighted combination of communication and radar metrics. Unlike \cite{hu2022low}, we optimize the multi-user interference, under a chirp similarity constraint for the radar performance, subject to \ac{PAPR} constraints and under an available power budget. Furthermore, the work in \cite{wang2010optimized} uses convex optimization tools to derive an iterative clipping and filtering in the frequency domain. However, \cite{wang2010optimized} does not enforce any multi-user communication metrics and radar metrics, but instead focuses on an \ac{OFDM} symbol level to minimize its \ac{EVM}. Waveform design based on constant modulus was proposed in \cite{cui2013mimo} for \ac{MIMO} radar, where the sequential optimization algorithm was adopted as a solution to the proposed framework. Even though the approach in \cite{cui2013mimo} takes into account radar, as well as \ac{PAPR} considerations \cite{bazzi2022method}, the approach does not account for multi-user communications. Similarly, a space-time coded design was adopted in \cite{cui2016space} and a Dinkelbach-Type algorithm \cite{barros1996new} was used as a solver towards their approach. Similar to \cite{cui2013mimo}, the approach in \cite{cui2016space} lacks multi-user communication capabilities. From a \ac{DFRC} perspective, the constant modulus design in \cite{liu2018toward} addresses multi-user communication and radar problem through the so-called \ac{BnB} method \cite{tuy1998convex}, thus guaranteeing a constant modulus waveform. Even though the solution is attractive in terms of \ac{PAPR}, it lacks flexibility in yielding a desired \ac{PAPR}. In many applications, the \ac{IBO} of the \ac{HPA}, which is basically the operating point of the amplifier backed-off from the saturation level, is directly related to the nominal transmit \ac{PAPR}. Therefore, in applications where \ac{IBO} can have margin, the design in \cite{liu2018toward} does not seem to be well-suited as it does not utilize the available power margin. Moreover, phase coded designs for radar performance is studied in \cite{de2008design} in order to maximize the radar's detection probability, but neither the \ac{PAPR} nor the multi-communication performances were addressed.
% Contibutions

In this paper, we address \ac{ISAC} waveform design from a practical point of view. We first propose an \ac{ISAC} waveform optimization framework, capable of multi-user interference minimization, while guaranteeing a similarity constraint relative to a radar chirp waveform. In addition, we take into account a practical factor, which is the \ac{PAPR} of the transmit waveform. The proposed design is capable of tuning the \ac{PAPR} to a desired level, a desired feature in practical PHY layer architectures. In principle, the general optimization framework falls under the class of \ac{MOO}, due to the various objectives involved in the problem \cite{bjornson2014multiobjective}. After forming an equivalent non-convex optimization problem, we employ an \ac{ADMM} based solution to converge towards an \ac{ISAC} waveform with desired \ac{PAPR} and radar properties. Finally, we highlight the various benefits of the proposed waveform design and the capability of the \ac{ADMM}-based waveform design solution in both radar sensing and multi-user communications, we present extensive simulation results showing the potential and superiority of the proposed design, when compared to state-of-the-art designs.

The remainder of this work is structured as follows. Section \ref{sec:system-model} introduces the sensing and communication system models. The \ac{KPIs} used in this paper are given in Section \ref{sec:kpi}. Section \ref{sec:isac-framework} introduces the \ac{ISAC} waveform design framework and formally states the optimization problem addressed in this paper. In Section \ref{sec:isac-admm}, the iterative solution in solving the problem is derived. Simulation results are demonstrated in Section \ref{sec:simulation-results}. Finally, we conclude the paper in Section \ref{sec:conclusions}.

\textbf{Notation}: Upper-case and lower-case boldface letters denote matrices and vectors, respectively. $(.)^T$, $(.)^*$ and $(.)^H$ represent the transpose, the conjugate and the transpose-conjugate operators. The statistical expectation is $\mathbb{E}\lbrace . \rbrace$. For any complex number $z \in \mathbb{C}$, the magnitude is $\vert z \vert$, its real part is $\Real(z)$, and  its imaginary part is $\Imag(z)$. For compactness, we denote the $i^{th}$ row of matrix $\mathbf{A}$ as $\mathbf{A}_{i}$. The Frobenius norm of matrix $\mathbf{X}$ is $\Vert \mathbf{X} \Vert$. The matrix $\mathbf{I}_N$ is the identity matrix of size $N \times N$. In particular, $\ve$ takes an $N\times M$ matrix $\mathbf{X}$ as input and returns an $NM \times 1$ vector, by stacking the columns of $\mathbf{X}$. The inverse of a square matrix is $\mathbf{X}^{-1}$. We index the $(i,j)^{th}$ entry of matrix $\mathbf{A}$ as $\mathbf{A}_{i,j}$.   The all-ones vector of size $N$ is denoted as $\mathbf{1}_N$ and $\mathbf{o}_k$ is an all-zeros vector, except for its $k^{th}$ entry, which is set to $1$. The Kronecker product is denoted as $\otimes$. The value at the $m^{th}$ iteration of a quantity, say $x$, involved in an iterative-type algorithm is denoted as $x^{(m)}$. The zero-vector is $\mathbf{0}$.  Furthermore, the vectorization and unvectorization operators are denoted as $\ve$ and $\ve^{-1}$, respectively.

\section{System Model}
\label{sec:system-model}
\begin{figure}[!t]
\centering
\includegraphics[width=3.5in]{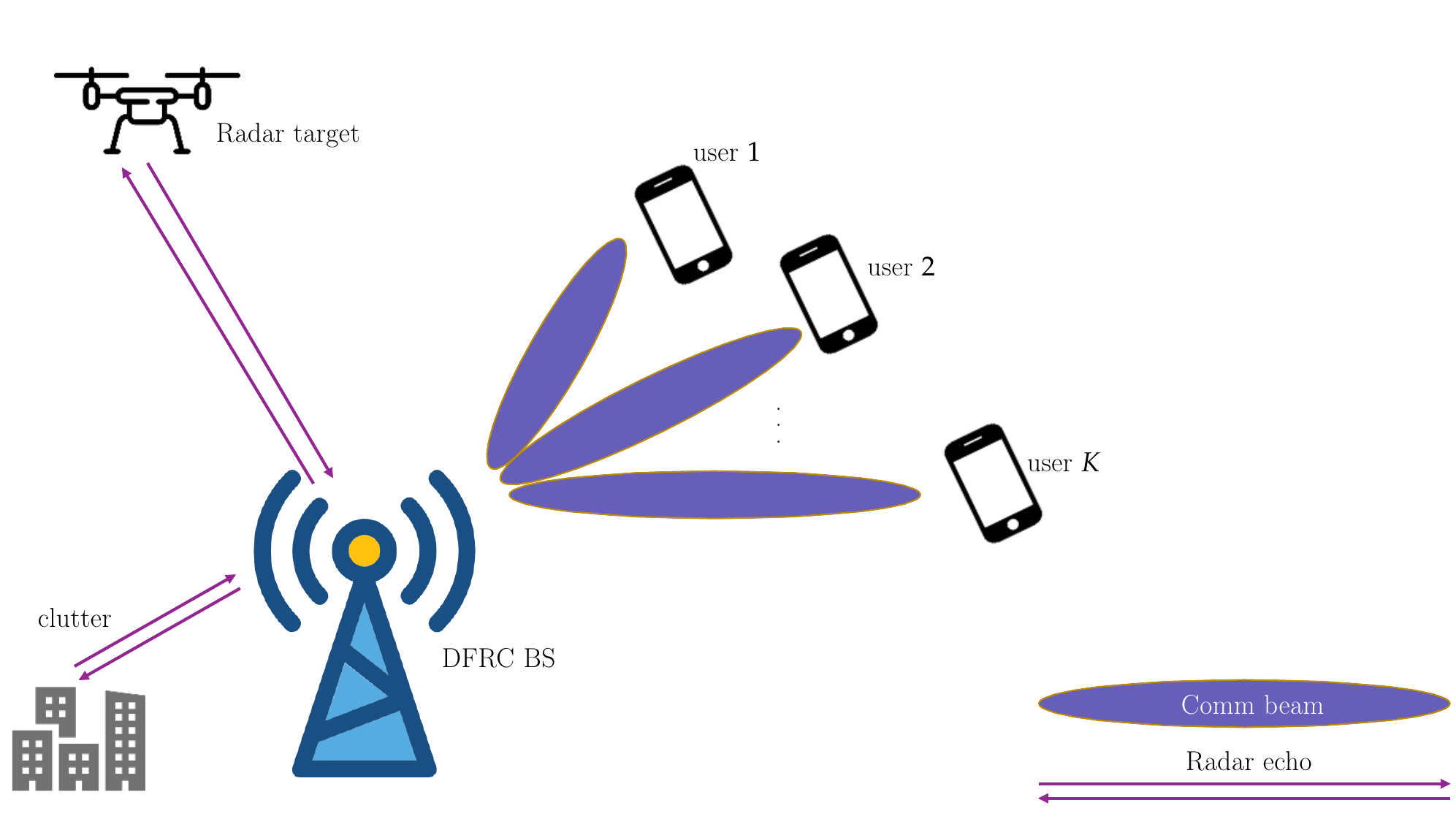}
\caption{An \ac{ISAC} architecture, where the scenario consists of a target of interest, $K$ communication users, and clutter due to scatterers in the environment.}
\label{fig:fig_1}
\end{figure}
\label{sec:system-model}
	Let us consider a colocated mono-static \ac{DFRC} \ac{BS} equipped with $N$ antennas in an arbitrary fashion. Also, consider a planar wave-front in signal propagation. The array steering vector of the antenna array at the \ac{DFRC} \ac{BS} is denoted by $\mathbf{a}(\theta)$ where $\theta$ is the \ac{AoA} of the wavefront. An example is the \ac{ULA} configuration, where
	\begin{equation}
		\mathbf{a}(\theta) 
		=
		\begin{bmatrix}
			1, & e^{-j \frac{2\pi d }{\lambda} \sin(\theta)}, & \ldots & e^{-j(N-1) \frac{2\pi d }{\lambda} \sin(\theta)} 
		\end{bmatrix}^T,
	\end{equation}
	where $d$ is the inter-element spacing and $\lambda$ is the wavelength. Communication users are considered to be located at random positions, whereas the target is supposed to be at a given angle $\theta_0$ from the \ac{DFRC} base station. Fig. \ref{fig:fig_1} depicts a \ac{DFRC} base station broadcasting the same \ac{ISAC} signal, hereby denoted as $\mathbf{X} \in \mathbb{C}^{N \times L}$ (where $L$ is the number of time samples), towards $K$ communication users and an intended target of interest.

%	 comprised of a target of interest, $K$ single-antenna communication users, and a \ac{DFRC} base station. 
\subsection{Sensing Echo Model}
The \ac{ISAC} \ac{DFRC} \ac{BS} is capable of sensing the desired echo
signal, which refers to its own echo signal reflected off intended targets.  Moreover, the echo signal encompasses the echo signal reflected by scatterers, referred to as environment clutter, and the echo signal reflected by targets of interest. Then, the received signal at the \ac{DFRC} \ac{BS} is as follows
\begin{equation}
	\label{eq:yr}
	\mathbf{Y}_r = \gamma_0 \mathbf{a}(\theta_0)\mathbf{a}^T(\theta_0)\mathbf{X} +  \sum\limits_{c = 1}^C \gamma_c \mathbf{a}(\theta_c)\mathbf{a}^T(\theta_c)\mathbf{X} + \mathbf{Z}_r \in \mathbb{C}^{N \times L}.
\end{equation}
The \ac{LoS} component towards the target of interest is captured in $\gamma_0 \mathbf{a}(\theta_0)\mathbf{a}^T(\theta_0)\mathbf{X}$, where $\gamma_0$ is the reflection coefficient containing attenuation of the \ac{LoS} component, as well as delay-doppler information within its phase. The \ac{AoA} of the target is $\theta_0$. Moreover, we consider $C$ clutter components, where the contribution of the $c^{th}$ scatterer is expressed by the term $\gamma_c \mathbf{a}(\theta_c)\mathbf{a}^T(\theta_c)\mathbf{X}$. Furthermore, due to the colocated mono-static architecture, the shared crystal clock at the \ac{DFRC} \ac{BS} leads to perfect synchronization between its \ac{UL}/\ac{DL} sense, therefore, the \ac{CFO} and the timing offset are zero \cite{zhang2021overview}. Also, we assume that the above reception, sampling and signal processing occur during a time interval termed the \ac{CPI} \cite{bazzi2022outage} which is an interval where sensing parameters are constant. Last but not least, the matrix $\mathbf{Z}_r$ reflects zero-mean background noise at the radar sub-system, where each column is assumed to be independent from the other.
\subsection{Communication \ac{DL} Model}
Using the same transmission, to leverage system resource, a single transmission in the \ac{DL} sense, can be expressed as 
\begin{equation}
	\label{eq:y=Hxpz}
	\mathbf{Y}_c = \mathbf{H}\mathbf{X} + \mathbf{Z}_c \in \mathbb{C}^{K \times L},
\end{equation}
where the $k^{th}$ row of $\mathbf{Y}_c$ is the received vector at the $k^{th}$ communication user. The channel matrix is given by $\mathbf{H} = \begin{bmatrix} \mathbf{h}_1 & \mathbf{h}_2 & \ldots & \mathbf{h}_K \end{bmatrix}^T \in \mathbb{C}^{K \times N}$, and is flat Rayleigh type fading, assumed to be constant during one transmission. Additionally, the \ac{BS} assumed complete channel knowledge $\mathbf{H}$. Finally, the vector $\mathbf{Z}_c \in \mathbb{C}^{K \times L}$ is background noise, where each column is white Gaussian i.i.d with zero mean and a multiple of identity covariance matrix as $\mathcal{N}(0,\sigma_c^2 \mathbf{I}_K)$.
\section{Key Performance Indicators}
\label{sec:kpi}
It is critical to specify \ac{KPIs} pertinent to the current system model in order to develop an appropriate and reliable optimization problem that attempts to solve \ac{ISAC} difficulties. We can rewrite the received signal as equation \eqref{eq:y=Hxpz} based on
\begin{equation}
	\label{eq:received-signal-2}
	\mathbf{Y}_c = \mathbf{S} + \underbrace{\mathbf{H}\mathbf{X} - \mathbf{S}}_{\text{MUI}} + \mathbf{Z}_c ,
\end{equation}
where MUI is multi-user interference and $\mathbf{S} \in \mathbb{C}^{K \times L}$ is the desired signal carrying information symbols. Following \cite{mohammed2013per}, it can be shown that the $\SINR$ at the $k^{th}$ user can be represented as a function of $\MUI$ as 
\begin{equation}
	\SINR_k
	=
	\frac{\mathbb{E}(\vert \mathbf{S}_{k,\ell} \vert^2)}{\mathbb{E} \big( \Vert (\mathbf{H}\mathbf{X} - \mathbf{S})_{k,\ell} \Vert_F^2 \big) + \sigma_c^2},
\end{equation}
and so the rate of the $k^{th}$ user is $R_k = \log_2( 1 + \SINR_k )$, following Shannon-theory. The MUI reflects a communication \ac{KPI}. Another important \ac{KPI} is the \ac{PAPR} is a waveform metric, which is the ratio of peak values to average power of that waveform. For example, a constant waveform enjoys a $\PAPR$ equal to one. The $\PAPR$ of a discrete time signal of $NL$ samples is expressed as
\begin{equation}
	\label{eq:papr-definition}
	\PAPR(\mathbf{x})
	=
	\frac{\max\limits_{\ell = 1 \ldots NL}\vert \mathbf{x}(\ell) \vert^2}{\frac{1}{NL}\sum\nolimits_{\ell = 1}^{NL} \vert \mathbf{x}(\ell) \vert^2}.
\end{equation}
As for sensing \ac{KPI}, we adopt a similarity constraint between the designed waveform, and a desired radar waveform, such as a given chirp waveform. Let the reference radar waveform be denoted as $\mathbf{x}_0$, hence a similarity constraint forms a sphere centered at the desired waveform $\mathbf{x}_0$ with radius $\epsilon$, 
\begin{equation}
	\mathbf{x} \in \mathcal{B}_{\epsilon}(\mathbf{x}_0) = \lbrace \mathbf{x} , \Vert \mathbf{x} - \mathbf{x}_0 \Vert^2 \leq \epsilon^2 \rbrace.
\end{equation}
In short, the MUI, \ac{PAPR} and the similarity constraints will serve as three \ac{KPIs} to the formulation of our \ac{ISAC} problem.

\section{\ac{ISAC} Waveform Design Framework}
\label{sec:isac-framework}
We formulate an optimization framework intended to maximize the total achievable rate of communication users, with sensing radar similarities and tunable \ac{PAPR}. It is worth noting that, under a unit-energy constellation, maximizing $\SINR_k$ is equivalent of minimizing $\mathbf{\MUI}_k$. We propose the following \ac{MOO} problem, where 
\begin{equation}
 \label{eq:MOO}
\begin{aligned}
(\mathcal{P}_{\tt{MOO}}):
\begin{cases}
\min\limits_{\lbrace \mathbf{x} \rbrace}&  \begin{bmatrix}
	\Vert  \mathbf{H}\mathbf{X} - \mathbf{S} \Vert^2_F  , & \Vert \mathbf{x} - \mathbf{x}_0 \Vert^2, & \PAPR(\mathbf{x})
\end{bmatrix} \\
\textrm{s.t.}
 & \Vert \mathbf{x} \Vert_F^2 = 1.\\ 
\end{cases}
\end{aligned}
\end{equation}
where $\mathbf{x} = \ve(\mathbf{X})$. Note that in the above \ac{MOO}, the aim is to find a waveform $\pmb{x}$ that joint minimizes the MUI, the radar similarity constraint and the \ac{PAPR}. Said differently, we would like to obtain a waveform that is close to both radar and communication metrics, with a low \ac{PAPR}. However, \ac{MOO}s are usually complex to solve in real time \cite{bjornson2014multiobjective}. As an alternative, we propose to minimize the MUI term, while bounding the radar similarity constraint and the \ac{PAPR}. To this end, we have the following alternate optimization problem,
\begin{equation}
 \label{eq:problem1}
\begin{aligned}
(\mathcal{P}):
\begin{cases}
\min\limits_{\lbrace \mathbf{x} \rbrace}&  \Vert  \mathbf{H}\mathbf{X} - \mathbf{S} \Vert^2_F \\
\textrm{s.t.}
 &  \PAPR(\mathbf{x}) \leq \eta \\ 
 & \mathbf{x} \in \mathcal{B}_{\epsilon}(\mathbf{x}_0) ,\\ 
  &  \Vert \mathbf{x} \Vert_F^2 = 1, \\
\end{cases}
\end{aligned}
\end{equation}
The problem is \eqref{eq:problem1} is proposed as an alternative to the \ac{MOO} in \eqref{eq:MOO}. The connection between \eqref{eq:problem1} and \eqref{eq:MOO} is that instead of jointly minimizing multi-user interference, radar chirp similarity, as well as the \ac{PAPR} in \eqref{eq:MOO}, we only choose to minimize the multi-user interference, when the similarity and \ac{PAPR} fall below certain pre-defined thresholds, controlled by $\epsilon$ and $\eta$.
 In \cite{bazzi2022integrated}, we show the equivalence of the problem $(\mathcal{P})$ to the following optimization problem
\begin{equation}
 \label{eq:problem3}
\begin{aligned}
(\mathcal{P}'):
\begin{cases}
\min\limits_{\lbrace \mathbf{x} \rbrace}&  \Vert  \mathbf{x} - \mathbf{x}_{\comm} \Vert^2_F \\
\textrm{s.t.}
 & \mathbf{x} \in \mathcal{B}_{\epsilon}(\mathbf{x}_0), \\ 
 &  \Vert \mathbf{x} \Vert_F^2 = 1, \\ 
 & \mathbf{x}^H \mathbf{F}_n \mathbf{x} \leq \frac{\eta}{NL} , \quad \forall n \\
\end{cases}
\end{aligned}
\end{equation}
where $\mathbf{x}_{\comm} = \ve(\mathbf{H}^H(\mathbf{H}\mathbf{H}^H)^{-1}\mathbf{S})$. Furthermore, $\mathbf{F}_n$ is a matrix of all-zeros, except for $1$, which is positioned at its $n^{th}$ diagonal entry. 
 Note the equivalence between the PAPR constraint in \eqref{eq:problem3} and that in \eqref{eq:papr-definition}, which is verified by designing an \ac{ISAC} waveform $\mathbf{x}$ under a norm-constraint, i.e. $\sum\nolimits_{\ell = 1}^{NL} \vert \mathbf{x}(\ell) \vert^2 = \Vert \mathbf{x} \Vert_F^2 = 1$, which when plugged in \eqref{eq:papr-definition} gives
 \begin{equation}
 	\max\limits_{\ell = 1 \ldots NL}\vert \mathbf{x}(\ell) \vert^2 \leq \frac{\eta}{NL},
 \end{equation}
 which is satisfied when each and every element  $\vert \mathbf{x}(\ell) \vert^2$ is upper-bounded by $\frac{\eta}{NL}$, hence the last constraint in \eqref{eq:problem3}.
 It can be easily noted that $(\mathcal{P}')$ is non-convex due to the norm-equality constraint. Indeed, $(\mathcal{P}')$ has several nice implications; the cost function could be interpreted as a similarity constraint between the waveform variable and a communication waveform, which is to be minimized. On the other hand, the first constraint is a radar similarity constraint, which is controlled by input $\epsilon$. The last two constraints are \ac{PAPR} constraints. In the following section, we describe an \ac{ADMM} procedure to solve $(\mathcal{P}')$ in an efficient way.

\section{\ac{ISAC} Design using \ac{ADMM}}
\label{sec:isac-admm}
The \ac{ADMM} method has emerged as a powerful technique for large-scale structured optimization problems with applications in different fields, such as distributed and scalable processing \cite{erseghe2012distributed} and image processing \cite{minaee2019admm}. \ac{ADMM} was initially devised as an iterative method for solving convex minimization problems by means of parallelization. However, it can also be applied in non-convex smooth optimization problems, such as group sparse problems \cite{6638818}. Before tackling the augmented Lagrangian problem, we introduce auxiliary variables to the problem at hand as, thus problem $(\mathcal{P}')$ is converted as 
\begin{equation}
 \label{eq:final-problem}
\begin{aligned}
(\bar{\mathcal{P}}'):
\begin{cases}
\min\limits_{\lbrace \bar{\mathbf{x}} \rbrace}&  \Vert   \bar{\mathbf{x}} -  \bar{\mathbf{x}}_{\comm} \Vert^2_F \\
\textrm{s.t.}
  & \mathbf{\alpha}  = \bar{\mathbf{x}}, \quad \mathbf{\alpha}^T \mathbf{\alpha} = 1,  \\
  & \mathbf{\beta} = \bar{\mathbf{x}} - \mathbf{x}_0, \quad \mathbf{\beta}^T \mathbf{\beta} \leq \epsilon^2 ,   \\
  & \mathbf{\gamma}_n  = \bar{\mathbf{F}}_n  \bar{\mathbf{x}} , \quad  \mathbf{\gamma}_n^T \mathbf{\gamma}_n  \leq \frac{\eta}{NL} , \quad \forall n \\
\end{cases}
\end{aligned}
\end{equation}
Note that in the above steps, we have also converted the problem to a real-valued one, i.e. $\bar{\mathbf{x}} = \begin{bmatrix} \Real(\mathbf{x})^T  &  \Imag(\mathbf{x})^T \end{bmatrix}^T$. Likewise, the same conversion has been applied to $\bar{\mathbf{x}}_{\comm}$ and $\bar{\mathbf{x}}_0$. Also, $\bar{\mathbf{F}}_n \in \mathbb{R} ^{2NL \times 2NL}$ is an all-zero matrix except its $n^{th}$ and $(NL+n)^{th}$ entries that are set to $1$. Furthermore, given a penalty parameter $\rho > 0$, the augmented Lagrangian of the problem $(\bar{\mathcal{P}}')$ could be expressed as 
\begin{equation}
\begin{split}
	&\mathcal{L}_{\rho}(\bar{\mathbf{x}},\mathbf{\alpha},\mathbf{\beta},\mathbf{\gamma}_n,\mathbf{u},\mathbf{v},\mathbf{w}_n) \\
	&=
	\Vert   \bar{\mathbf{x}} -  \bar{\mathbf{x}}_{\comm} \Vert^2_F
	+ \mathbf{u}^T ( \bar{\mathbf{x}} - \mathbf{\alpha} ) + \mathbf{v}^T ( \bar{\mathbf{x}} - \bar{\mathbf{x}}_0 - \mathbf{\beta}) \\
	&+  \sum\limits_{n=1}^{NL} \mathbf{w}_n^T (\bar{\mathbf{F}}_n  \bar{\mathbf{x}}  - \mathbf{\gamma}_n) + \frac{\rho}{2} \Vert \bar{\mathbf{x}} - \mathbf{\alpha} \Vert^2 + \frac{\rho}{2} \Vert \bar{\mathbf{x}} - \bar{\mathbf{x}}_0 - \mathbf{\beta} \Vert^2 \\
	&+  \frac{\rho}{2} \sum\nolimits_{n=1}^{NL} \Vert \bar{\mathbf{F}}_n  \bar{\mathbf{x}}  - \mathbf{\gamma}_n \Vert^2, 
\end{split}
\end{equation}
The variables are updated in following an iterative alternating pattern and over the regions of interest, namely
\begin{subequations}
	\begin{equation}
		\mathcal{C}_{\alpha} = \lbrace \mathbf{\alpha} , \quad \mathbf{\alpha}^T \mathbf{\alpha} = 1 	 \rbrace , 
	\end{equation}
	\begin{equation}
		\mathcal{C}_{\beta} = \lbrace \mathbf{\beta} , \quad  \mathbf{\beta}^T \mathbf{\beta} \leq \epsilon^2  \rbrace , 
	\end{equation}
	\begin{equation}
		\mathcal{C}_{\gamma} = \lbrace \mathbf{\gamma} , \quad \mathbf{\gamma}^T \mathbf{\gamma} \leq \frac{\eta}{NL} \rbrace , 
	\end{equation}
\end{subequations}
In particular, at iteration $m$, the vector $\bar{\mathbf{x}}^{(m+1)}$ minimizes $\mathcal{L}_{\rho}(\bar{\mathbf{x}},\mathbf{\alpha}^{(m)},\mathbf{\beta}^{(m)},\mathbf{\gamma}_n^{(m)},\mathbf{u}^{(m)},\mathbf{v}^{(m)},\mathbf{w}_n^{(m)})$. In the second update, given $\bar{\mathbf{x}}^{(m+1)}$, the vector $\mathbf{\alpha}^{(m+1)}$ minimizes $\mathcal{L}_{\rho}(\bar{\mathbf{x}}^{(m+1)},\mathbf{\alpha},\mathbf{\beta}^{(m)},\mathbf{\gamma}_n^{(m)},\mathbf{u}^{(m)},\mathbf{v}^{(m)},\mathbf{w}_n^{(m)})$. Then, given $\bar{\mathbf{x}}^{(m+1)},\mathbf{\alpha}^{(m+1)}$, $\mathbf{\beta}^{(m+1)}$ minimizes $\mathcal{L}_{\rho}(\bar{\mathbf{x}}^{(m+1)},\mathbf{\alpha}^{(m+1)},\mathbf{\beta},\mathbf{\gamma}_n^{(m)},\mathbf{u}^{(m)},\mathbf{v}^{(m)},\mathbf{w}_n^{(m)})$. The last update utilizes all the current values, i.e. $\bar{\mathbf{x}}^{(m+1)},\mathbf{\alpha}^{(m+1)},\mathbf{\beta}^{(m+1)}$ to update $\mathbf{\gamma}_n^{(m+1)}$ by minimizing $\mathcal{L}_{\rho}(\bar{\mathbf{x}}^{(m+1)},\mathbf{\alpha}^{(m+1)},\mathbf{\beta}^{(m+1)},\mathbf{\gamma}_n,\mathbf{u}^{(m)},\mathbf{v}^{(m)},\mathbf{w}_n^{(m)})$. Last but not least, before proceeding to iteration $m+2$, the auxiliary variables are updated in a linear fashion as follows,
\begin{align}
	\mathbf{u}^{(m+1)} &= \mathbf{u}^{(m)} + \rho(\bar{\mathbf{x}}^{(m+1)} - \mathbf{\alpha}^{(m+1)}) \label{eq:update-u}, \\
	\mathbf{v}^{(m+1)} &= \mathbf{v}^{(m)} + \rho(\bar{\mathbf{x}}^{(m+1)} - \mathbf{x}_0 - \mathbf{\beta}^{(m+1)}) \label{eq:update-v}, \\
	\mathbf{w}_n^{(m+1)} &= \mathbf{w}_n^{(m)} + \rho(\bar{\mathbf{F}}_n  \bar{\mathbf{x}}^{(m+1)} - \mathbf{\gamma}_n^{(m+1)}) \label{eq:update-w}.	
\end{align}
\begin{algorithm}
\caption{{\ac{ADMM}-based \ac{DFRC} Waveform Design}}\label{alg:cap}
\begin{algorithmic}
\State {\textbf{INPUT}} $\mathbf{x}_0, \mathbf{H}, \mathbf{S}$
\State {\textbf{INIT}} $\mathbf{\alpha}^{(0)} = \mathbf{0},\mathbf{\beta}^{(0)} = \mathbf{0},\mathbf{\gamma}_1^{(0)} =\mathbf{\gamma}_2^{(0)} =\ldots =\mathbf{\gamma}_{NL}^{(0)} = \mathbf{0}$
\State {\textbf{INIT}} $\mathbf{u}^{(0)} = \mathbf{0},\mathbf{v}^{(0)} = \mathbf{0},\mathbf{w}_1^{(0)} =\mathbf{w}_2^{(0)} =\ldots =\mathbf{w}_{NL}^{(0)} = \mathbf{0}$
\State {\textbf{SET}} $m \gets 0$, $\mathbf{x}_{\comm} = \ve(\mathbf{H}^H(\mathbf{H}\mathbf{H}^H)^{-1}\mathbf{S})$
\State {\textbf{FORM}} $\bar{\mathbf{x}}_{\comm} = \begin{bmatrix}
	\Real({\mathbf{x}}_{\comm})^T & \Imag({\mathbf{x}}_{\comm})^T
\end{bmatrix}^T$
\State {\textbf{FORM}} $\bar{\mathbf{x}}_{0} = \begin{bmatrix}
	\Real({\mathbf{x}}_{0})^T & \Imag({\mathbf{x}}_{0})^T
\end{bmatrix}^T$
\While{$m < M_{\tt{iter}}$}
\State {Update} $\mathbf{x}^{(m+1)}$ {using equation} \eqref{eq:x-update}
\State {Update} $\mathbf{\alpha}^{(m+1)}$ {using equation }\eqref{eq:alpha-update}
\State {Update} $\mathbf{\beta}^{(m+1)}$ {using equation }\eqref{eq:beta-update} 
\State {Update} $\mathbf{\gamma}_1^{(m+1)} \ldots \mathbf{\gamma}_{NL}^{(m+1)}$ {via equation} \eqref{eq:gamma-update}
\State {Update} $\mathbf{u}^{(m+1)} \gets \mathbf{u}^{(m)} + \rho(\bar{\mathbf{x}}^{(m+1)} - \mathbf{\alpha}^{(m+1)}) $
\State {Update} $\mathbf{v}^{(m+1)} \gets \mathbf{v}^{(m)} + \rho(\bar{\mathbf{x}}^{(m+1)} - \mathbf{x}_0 - \mathbf{\beta}^{(m+1)})$
\State {Update} $\mathbf{w}_n^{(m+1)} \gets \mathbf{w}_n^{(m)} + \rho(\bar{\mathbf{F}}_n  \bar{\mathbf{x}}^{(m+1)} - \mathbf{\gamma}_n^{(m+1)})$
\State $m \gets m + 1$
\EndWhile
\State 
\Return $\mathbf{x}^{(M_{\tt{iter}})}$
\end{algorithmic}
\end{algorithm}

To find $\bar{\mathbf{x}}^{(m+1)}$, we set the gradient of $\mathcal{L}_{\rho}(\bar{\mathbf{x}},\mathbf{\alpha}^{(m)},\mathbf{\beta}^{(m)},\mathbf{\gamma}_n^{(m)},\mathbf{u}^{(m)},\mathbf{v}^{(m)},\mathbf{w}_n^{(m)})$ with respect to $\bar{\mathbf{x}}$ to zero,
\begin{equation}
\label{eq:gradient-at-x}
	\nabla_{\bar{\mathbf{x}}}
	\Big[
	\mathcal{L}_{\rho}(\bar{\mathbf{x}},\mathbf{\alpha}^{(m)},\mathbf{\beta}^{(m)},\mathbf{\gamma}_n^{(m)},\mathbf{u}^{(m)},\mathbf{v}^{(m)},\mathbf{w}_n^{(m)})
	\Big]_{\bar{\mathbf{x}} = \bar{\mathbf{x}}^{(m+1)} } 
	=
	\mathbf{0},
\end{equation}
which gives.
\begin{equation}
\label{eq:x-update}
\begin{split}
	\bar{\mathbf{x}}^{(m+1)} 
	&=
	\frac{1}{2+3\rho}
	\Bigg(
	2 \bar{\mathbf{x}}_{\comm} 
	- \mathbf{u}^{(m)} 
	- \mathbf{v}^{(m)}
	- \sum\nolimits_{n=1}^{NL}\bar{\mathbf{F}}_n \mathbf{w}_n^{(m)} \\ &
	+ \rho \mathbf{\alpha}^{(m)} + \rho (\bar{\mathbf{x}}_0 + \mathbf{\beta}^{(m)} )  
	+ \rho \sum\nolimits_{n=1}^{NL} \bar{\mathbf{F}}_n \mathbf{\gamma}_n^{(m)}
	\Bigg).
\end{split}
\end{equation}
For details in derivations, the reader is referred to \cite{bazzi2022integrated}. Following similar steps, the update equations read
\begin{equation}
\label{eq:alpha-update}
	\mathbf{\alpha}^{(m+1)}
	=
	\frac{\bar{\mathbf{x}}^{(m+1)} + \frac{1}{\rho} \mathbf{u}^{(m)} }{\Vert \bar{\mathbf{x}}^{(m+1)} + \frac{1}{\rho} \mathbf{u}^{(m)}  \Vert },
\end{equation}

\begin{equation}
 \label{eq:beta-update}
	\mathbf{\beta}^{(m+1)}
	=
\begin{cases}
\bar{\mathbf{x}}^{(m+1)} - \bar{\mathbf{x}}_0 + \frac{1}{\rho} \mathbf{v}^{(m)} , & \text{if  } \in \mathcal{C}_\beta \\
		\epsilon
	\frac{\bar{\mathbf{x}}^{(m+1)} - \bar{\mathbf{x}}_0 + \frac{1}{\rho} \mathbf{v}^{(m)} }{\Vert \bar{\mathbf{x}}^{(m+1)} - \bar{\mathbf{x}}_0  + \frac{1}{\rho} \mathbf{v}^{(m)}  \Vert } , & \text{otherwise} .
\end{cases}
\end{equation}

\begin{equation}
\label{eq:gamma-update}
	\mathbf{\gamma}_n^{(m+1)}  
	=
	\begin{cases}
	\bar{\mathbf{F}}_n  \bar{\mathbf{x}}^{(m+1)} + \frac{1}{\rho} \mathbf{w}_n^{(m)}, & \text{if }  \in \mathcal{C}_\gamma \\
		\sqrt{\frac{\eta}{NL}}\frac{ \bar{\mathbf{F}}_n  \bar{\mathbf{x}}^{(m+1)} + \frac{1}{\rho} \mathbf{w}_n^{(m)} }{\Vert \bar{\mathbf{F}}_n  \bar{\mathbf{x}}^{(m+1)} + \frac{1}{\rho} \mathbf{w}_n^{(m)} \Vert},
		& \text{otherwise}.
	\end{cases}
\end{equation}
A summary of the proposed \ac{ADMM}-based \ac{DFRC} waveform design is summarized in {\tt{\textbf{{Algorithm \ref{alg:cap}}}}}. Due to space restrictions in this manuscripts, the convergence analysis and complexity analysis of the proposed method can be found in our work in \cite{bazzi2022integrated}. Also, note that this work assumes perfect channel state information, which is not the case in practice. A newly proposed algorithm is also found in \cite{bazzi2022integrated} to cope with imperfect channel state information arising from practical scenarios, such as estimation errors.

%This paper consider that the CSI is perfect, however, in practice, the CSI is always not perfect due to estimation error. So how the imperfect CSI affect the performance of the proposed method

\section{Simulation Results}
\label{sec:simulation-results}
This section uses a variety of simulation results to show the effectiveness and trade-offs of the proposed \ac{ISAC}-waveform design. The channel $\pmb{H}$ is drawn from a complex gaussian distribution and the desired constellation $\pmb{S}$ contains randomly selected \ac{QAM} symbols.

\begin{figure}[t]
	\centering
	\includegraphics[width=1\linewidth]{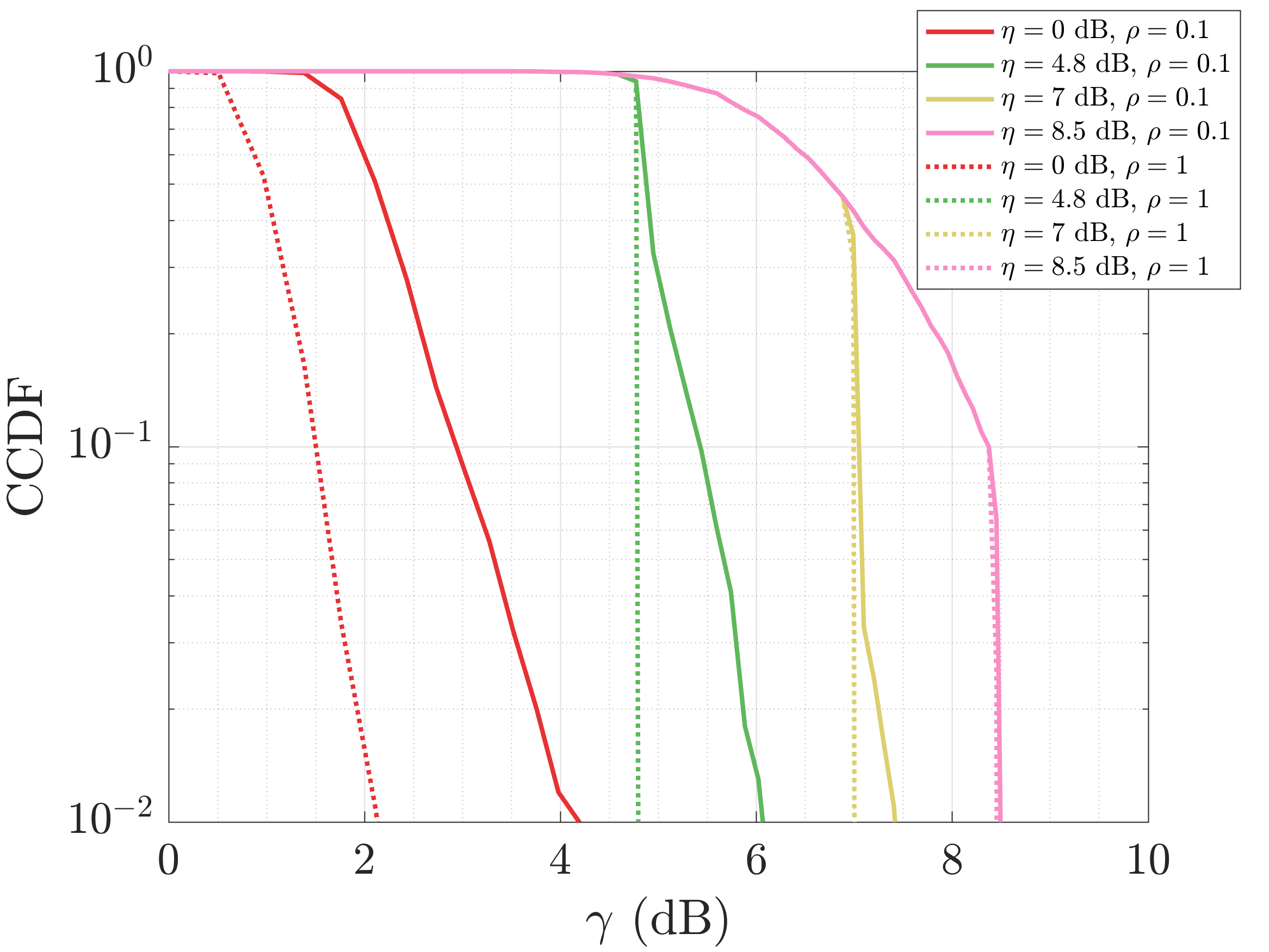}
	\caption{The CCDF of the generated waveform described in Algorithm 1 for different values of $\eta$ and $\rho$.}
	\label{fig:CCDF}
\end{figure}

\begin{figure}[t]
	\centering
	\includegraphics[width=1\linewidth]{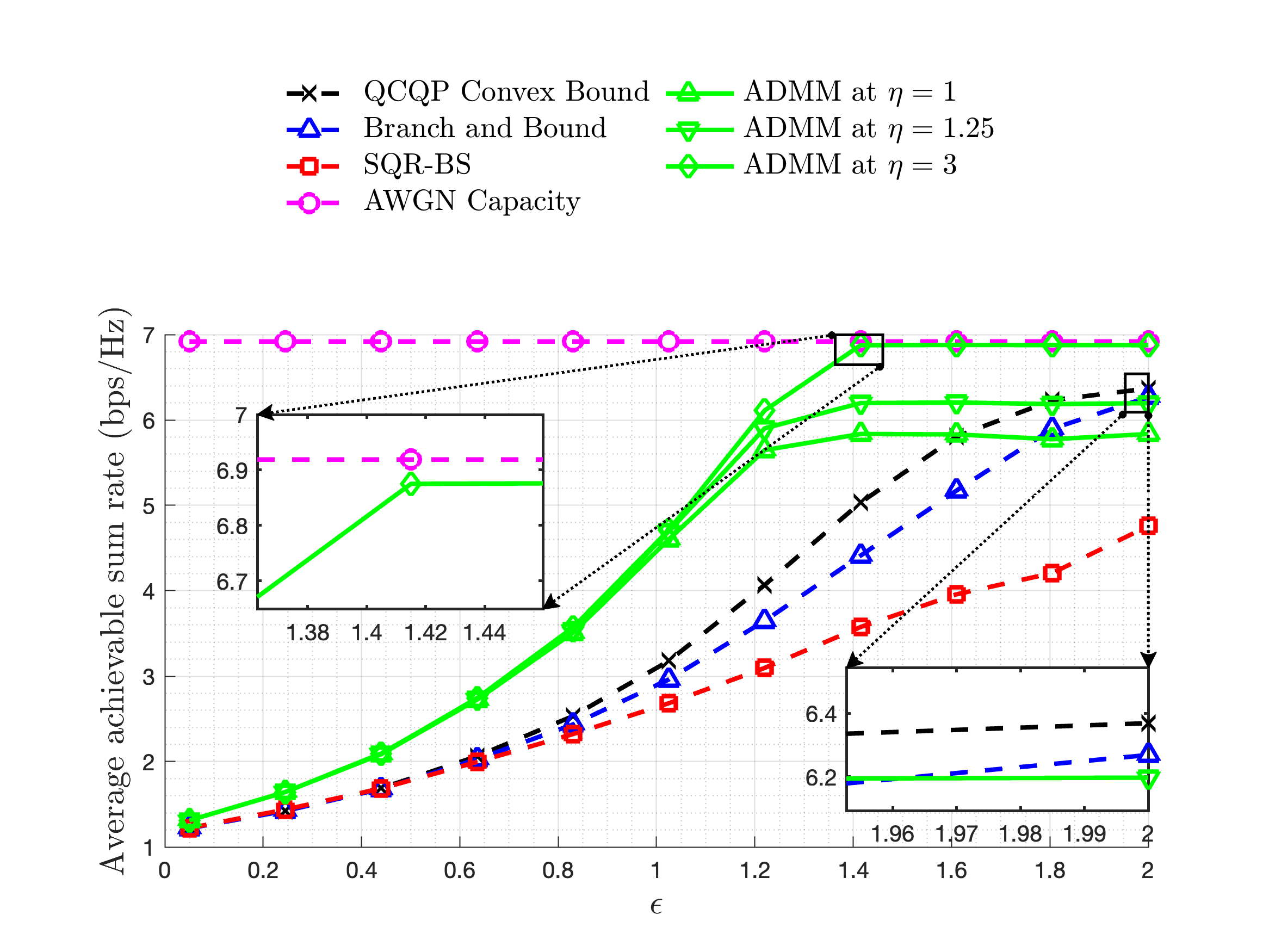}
	\caption{Trade-off between the communication average achievable sum- rate per user and similarity waveform radar measure at $\SNR=10$ dB for $N = 4$ antennas and $K = 2$ communication users. }
	\label{fig:SumRate}
\end{figure}

\begin{figure}[t]
	\centering
	\includegraphics[width=1\linewidth]{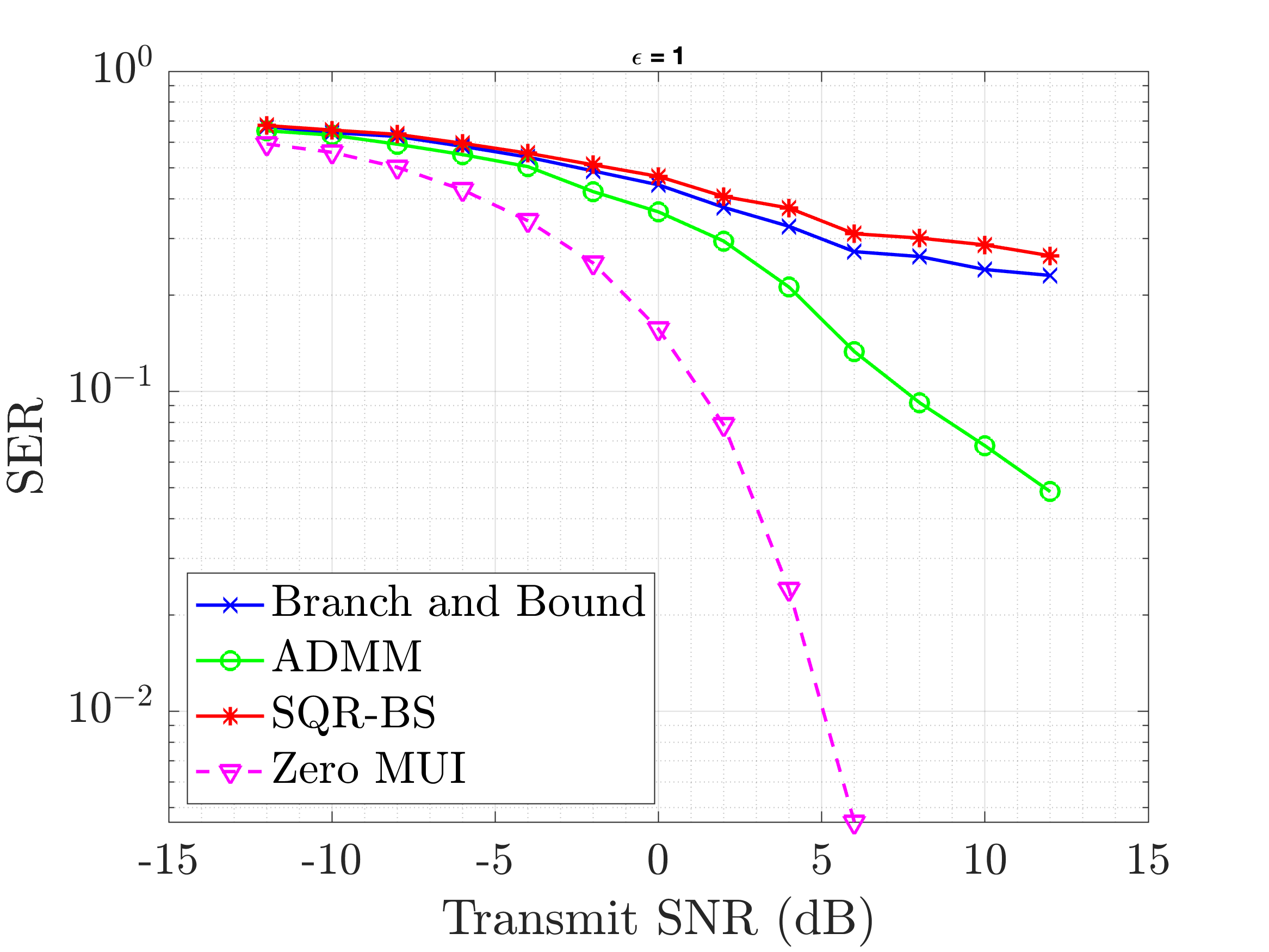}
	\caption{The symbol error rate performance at $\epsilon = 1$ for $N = 5$ antennas and $K = 2$ communication users. The constellation utilized is a QPSK.}
	\label{fig:SER}
\end{figure}
%% fig:CCDF
In Fig. \ref{fig:CCDF}, we plot the resulting \ac{CCDF} of the measured \ac{PAPR} for different values of $\rho,\eta$. Note that the \ac{CCDF} is computed as 
\begin{equation}
	\CCDF \triangleq \Pr(\PAPR(\mathbf{x}) > \gamma).
\end{equation} 
It is worth noting that increasing $\rho$ results in a steeper cutoff of the \ac{PAPR} \ac{CCDF} most notable for larger values of $\eta$. For example, fixing the \ac{CCDF} probability of $10^{-2}$ and a target \ac{PAPR} of $\eta = 0$dB, we see that for $\rho = 0.1$, the required $\gamma$ is $\gamma = 4.19$dB, whereas for $\rho = 1$, the $\gamma$ is decreased by $2$dB. On the other hand, tolerating a higher \ac{PAPR}, this gap is reduced. Indeed, for $\eta = 4.8$dB, we can observe that for $\rho = 1$, the requirement is already satisfied at a \ac{CCDF} probability of $10^{-2}$, as opposed to when $\rho = 0.1$, we see that $\gamma = 6.11$dB, which reflects a gap of about $1.31$dB. 
For a larger $\eta$, say $\eta = 7$dB, we see that the gap in $\gamma$ is further reduced to $0.4$dB between $\rho = 1$ and $\rho = 0.1$. Finally, this gap becomes negligible at $\eta = 8.5$dB.
% fig:SumRate
In Fig. \ref{fig:SumRate}, our goal is to investigate the trade-off between radar waveform similarity and communications sum-rate. The figure depicts the average achievable sum-rate as a function of $\epsilon$ for $\SNR = 10$dB, $N = 4$ and $K = 2$. Fig. \ref{fig:SumRate} delineates the average achievable sum-rate, as a function of $\epsilon$ at $\SNR = 10$dB, $N = 4$ transmit antennas and $K = 2$ communication users. Also, some benchmarks are employed such as the successive \ac{QCQP} refinement (SQR) binary search (SQR-BS) algorithm proposed in \cite{aldayel2016successive}, the \ac{BnB} method \cite{liu2018toward}, and the \ac{AWGN} capacity. For $\epsilon < 1.6$, the performance of the proposed \ac{ADMM} waveform design at $\eta = 1$ outperforms that of the QCQP convex bound and for $\epsilon < 1.8$ the \ac{ADMM} method outperforms the \ac{BnB} method. Note that all methods outperform the SQR-BS method, which agrees with the results in \cite{liu2018toward}. For a higher $\eta$, i.e. $\eta = 1.25$, we can observe that \ac{ADMM} outperforms \ac{BnB} for any $\epsilon$. Additionally, we see that the \ac{ADMM} achieves the \ac{AWGN} capacity performance at $\epsilon > 1.42$ for $\eta = 3$. The capability of \ac{PAPR} adjustment explains the gain of the proposed \ac{ISAC} \ac{ADMM}-based design, which has a direct effect on sum-rate.

% fig:SER
In Fig. \ref{fig:SER}, we now demonstrate the performance of communication for $\epsilon = 1$, which could be seen as "mid-way" between radar and communications. We are specifically interested with the \ac{SER} performances, as a function of transmit $\SNR$. We employ \ac{QPSK} with random symbols streamed towards $K = 2$ communication users, via $N = 5$ transmit antennas at the \ac{DFRC} \ac{BS}. Now, we compare the \ac{SER} of the proposed \ac{ADMM} method compared to \ac{BnB}, SQR-BS and a benchmark of zero MUI. For $\epsilon = 1$ and at a level of $10^{-1}$ \ac{SER}, we see that the proposed \ac{ISAC} waveform design based on \ac{ADMM} is about $6$dB away from the zero MUI performance. In addition, the proposed \ac{ADMM} design outperforms \ac{BnB} and SQR-BS by more than $7$dB at a level of $10^{-1}$ \ac{SER}. Therefore, this reveals the superiority of the proposed \ac{ADMM} waveform design.

\section{Conclusions}
\label{sec:conclusions}
In this paper, we have proposed an \ac{ISAC} waveform design with desired radar waveform similarity, carrying multi-user communication information, and with adjustable \ac{PAPR}. Indeed, this design is suitable for \ac{ISAC} applications, where the \ac{IBO} of the \ac{HPA} is an important factor. Additionally, the trade-off between radar and communications has been revealed through simulations. Moreover, simulation results unveil the superiority of the proposed \ac{ADMM}-based \ac{ISAC} waveform design, as compared to state-of-the-art \ac{ISAC} waveform.  
\bibliographystyle{IEEEtran}

\bibliography{refs}

%\begin{thebibliography}{00}
%\bibitem{b1} G. Eason, B. Noble, and I. N. Sneddon, ``On certain integrals of Lipschitz-Hankel type involving products of Bessel functions,'' Phil. Trans. Roy. Soc. London, vol. A247, pp. 529--551, April 1955.
%\bibitem{b2} J. Clerk Maxwell, A Treatise on Electricity and Magnetism, 3rd ed., vol. 2. Oxford: Clarendon, 1892, pp.68--73.
%\bibitem{b3} I. S. Jacobs and C. P. Bean, ``Fine particles, thin films and exchange anisotropy,'' in Magnetism, vol. III, G. T. Rado and H. Suhl, Eds. New York: Academic, 1963, pp. 271--350.
%\bibitem{b4} K. Elissa, ``Title of paper if known,'' unpublished.
%\bibitem{b5} R. Nicole, ``Title of paper with only first word capitalized,'' J. Name Stand. Abbrev., in press.
%\bibitem{b6} Y. Yorozu, M. Hirano, K. Oka, and Y. Tagawa, ``Electron spectroscopy studies on magneto-optical media and plastic substrate interface,'' IEEE Transl. J. Magn. Japan, vol. 2, pp. 740--741, August 1987 [Digests 9th Annual Conf. Magnetics Japan, p. 301, 1982].
%\bibitem{b7} M. Young, The Technical Writer's Handbook. Mill Valley, CA: University Science, 1989.
%\end{thebibliography}
\vspace{12pt}

\end{document}